\documentstyle[aps,prl,multicol,epsf,psfig,amssymb]{revtex}

\begin{document}

\tightenlines

\title{Phase Ordering and Onset of Collective Behavior in
Chaotic Coupled Map Lattices}


\author{Ana\"el Lema\^{\i}tre$^{(1,2)}$ and Hugues Chat\'e$^{(2,1)}$}
\address{$^{(1)}$LadHyX, Laboratoire d'Hydrodynamique, Ecole Polytechnique,
91128 Palaiseau, France\\
$^{(2)}$CEA, Service de Physique de l'Etat Condens\'e,
Centre d'Etudes de Saclay, 91191 Gif-sur-Yvette, France\\
}
\date{\today}

\maketitle
\begin{abstract}
The phase ordering properties of lattices of band-chaotic maps coupled 
diffusively with some coupling strength $g$ are studied in order to 
determine the limit value $g_{\rm e}$ beyond which multistability
disappears and non-trivial collective behavior is observed. 
The persistence of equivalent discrete spin variables and the
characteristic length of the patterns observed scale algebraically
with time during phase ordering. The associated exponents vary continuously
with $g$ but remain proportional to each other, with a ratio close to that
of the time-dependent Ginzburg-Landau equation.
The corresponding individual values 
seem to be recovered in the space-continuous limit.
\end{abstract}


\begin{multicols}{2}

One of the most remarkable features distinguishing 
 extensively-chaotic dynamical systems from most models studied in
out-of-equilibrium statistical physics is that they
generically exhibit  non-trivial collective behavior (NTCB), 
i.e. long-range order emerging out of local chaos, 
accompanied by the temporal evolution of 
spatially-averaged quantities \cite{GEN-NTCB,CA-NTCB,CML-NTCB}.
In particular, NTCB is easily observed on simple models of reaction-diffusion
systems such as coupled map lattices (CMLs) in which (chaotic)
nonlinear maps $S$ of real variables $X$ are coupled diffusively with some 
coupling strength $g$ \cite{CML-NTCB}.

NTCB is often claimed
to be a {\it macroscopic} attractor, well-defined in the infinite-size
limit and reached for almost every initial
condition,
provided the local coupling between sites is ``large enough''.
On the other hand, for small $g$ values, such as those corresponding
to the so-called
``anti-integrable'' limit which tries to extend zero-coupling behavior
to small, but finite coupling strengths, 
CMLs often exhibit multistability \cite{ANTI-MULTI}.
This is in particular the case if the local map shows banded chaos, 
because the interfaces
separating clusters of sites in the different bands can be pinned. This
multistability is ``extensive'':
the number of (chaotic) attractors
may then be argued to grow exponentially
with the system size, in opposition to NTCB for which this number is
small and size-independent.

In this Letter, we 
define and measure the limit coupling strength $g_{\rm e}$ separating 
the strong-coupling regime in which NTCB is observed from 
the weak-coupling, extensive multistability region.
Using  the discrete ``spin'' variables which can be defined whenever  the
one-body probability distribution functions (pdfs) 
of local (continuous) variables have  disjoint supports,
we study numerically 
the phase ordering process following uncorrelated initial conditions in cases
where the spin variables take only two values.
We find that the persistence probability $p(t)$ 
(i.e. the proportion of spins which have not changed sign up to time $t$) 
saturates in finite time
to strictly positive values in the weak coupling regime 
whereas it decays algebraically to zero when $g>g_{\rm e}$. 
The associated persistence exponent $\theta$ 
varies continuously with 
parameters, at odds with traditional models \cite{PERSIS}.
Moreover, data obtained on various two-dimensional CMLs
is best accounted for by a relation of the form $\theta \sim (g-g_{\rm e})^w$,
which we use to estimate $g_{\rm e}$.
We show further that this behavior is mostly due to the non-trivial 
scaling of the characteristic length $L(t)\sim t^\phi$ 
during the phase ordering
process. Indeed, $\phi \ne \frac{1}{2}$, the expected value for a 
scalar, non-conserved order parameter \cite{MODELA},
and is found to be proportional to $\theta$, with the exponent ratio
$\phi/\theta$ approximately taking the value known for the time-dependent
Ginzburg-Landau equation (TDGLE).
We also provide evidence that,
in the continuous-space limit, ``normal'' phase ordering
behavior is recovered.
Finally, we discuss the hierarchy of limit coupling values $g_{\rm e}^n$
which can be defined when the local map is unimodal and
shows $2^n$-band chaos, using 
recent results on renormalisation group (RG) ideas applied to CMLs 
\cite{CMLRG}.

Consider a  $d$-dimensional hypercubic lattices ${\cal L}$
of coupled identical maps $S_\mu$ acting on real variables
$(X_{\vec r})_{{\vec r} \in {\cal L}}$:
\begin{equation}
X_{\vec r}^{t+1} = (1-2dg) S_{\mu}(X_{\vec r}^t) + 
g \sum_{{\vec e} \in {\cal V}} S_{\mu}(X_{{\vec r}+{\vec e}}^t) \;,
\label{eq-cml}
\end{equation}
where ${\cal V}$ is the set of $2d$ nearest neighbors $\vec e$ of site 
$\vec 0$. 
We first present results obtained for the piecewise linear, odd, 
local map $S_{\mu}$ defined by:
\begin{equation}
S_{\mu}(X) = \left\{
\begin{array}{lll}
\mu X & {\rm if} & X \in [-1/3,1/3] \\
2\mu/3 - \mu X & {\rm if} & X \in [1/3,1] \\
-2\mu/3 - \mu X & {\rm if} & X \in [-1,-1/3]
\end{array} \right.
\label{eq-mhmap}
\end{equation}
which leaves the $I=[-1,1]$ interval invariant. (For $\mu=3$, this is the 
chaotic map introduced by Miller and Huse \cite{MH}.) For $\mu\in [-2,-1]$,
$S_\mu$ displays banded chaos, while for opposite $\mu$ values, these bands
become invariant subintervals of $I$. At $\mu=1.9$ in particular, $S_\mu$
possesses two symmetric such intervals 
$I^\pm=[\pm \mu (2-\mu)/3, \pm \mu/3]$,
separated by a finite gap. 
For any value of $g$, the support of the pdf of $X$ for the 
CML defined by (\ref{eq-cml}-\ref{eq-mhmap}) can be separated into two 
components thanks to the symmetry of the map.
This allows the unambiguous definition
of spin variables $\sigma_{\vec r}= {\rm sign}(X_{\vec r})$. The deterministic
nature of the system and the form of the coupling 
strictly forbids the nucleation of opposite-phase droplets in clusters:
the analog spin system is at zero temperature.

\begin{figure}
\narrowtext
\unitlength = 0.0011\textwidth
\begin{center}
\begin{picture}(200,200)(0,15)
\put(100,205){\makebox(0,0){\large (a)}}
\put(0,0){\makebox(200,200){\epsfxsize=180\unitlength
\epsffile{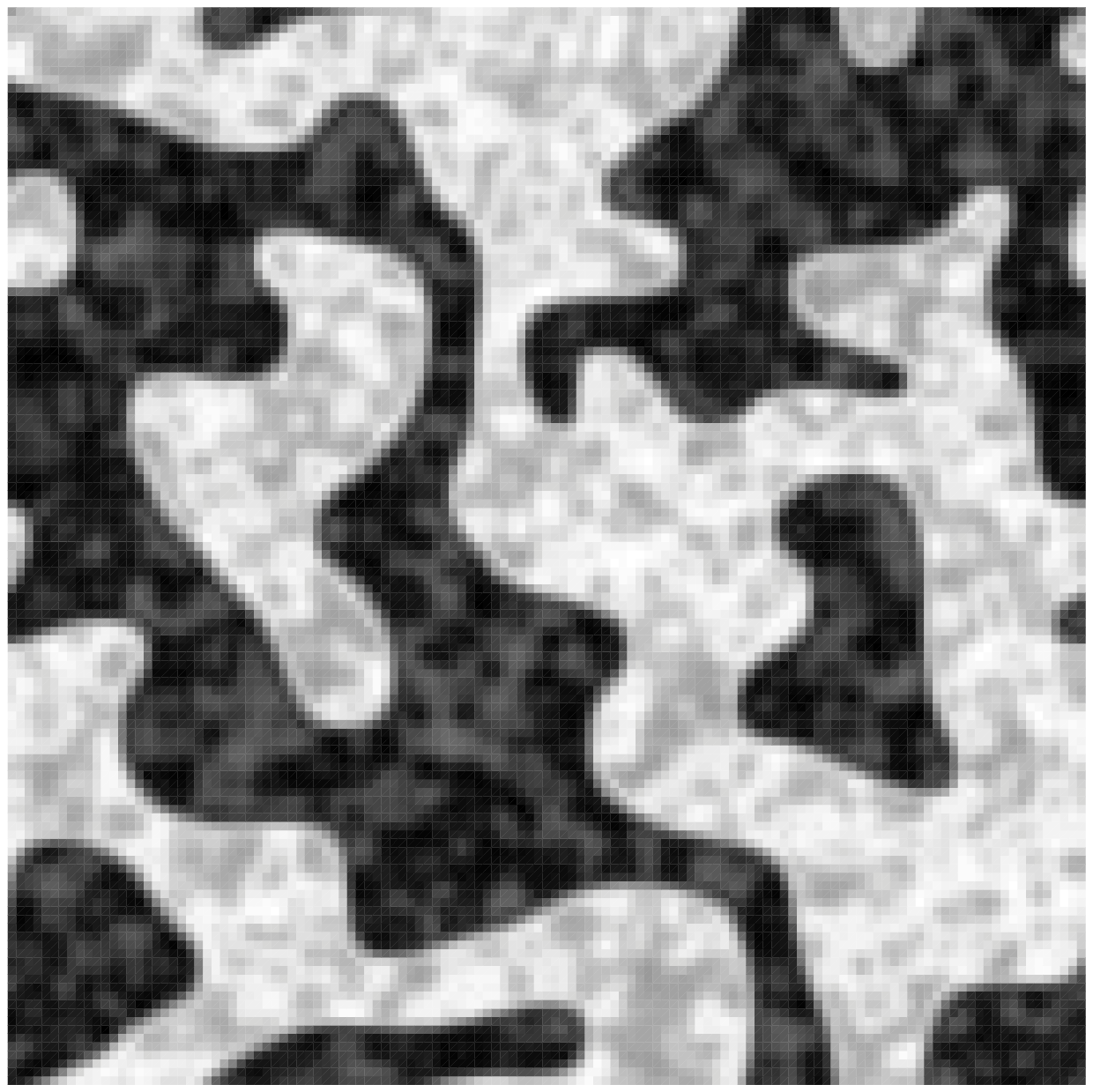}}}
\end{picture}
\hspace{10\unitlength}
\begin{picture}(200,200)(0,15)
\put(100,205){\makebox(0,0){\large (b)}}
\put(0,0){\makebox(200,200){\epsfxsize=180\unitlength
\epsffile{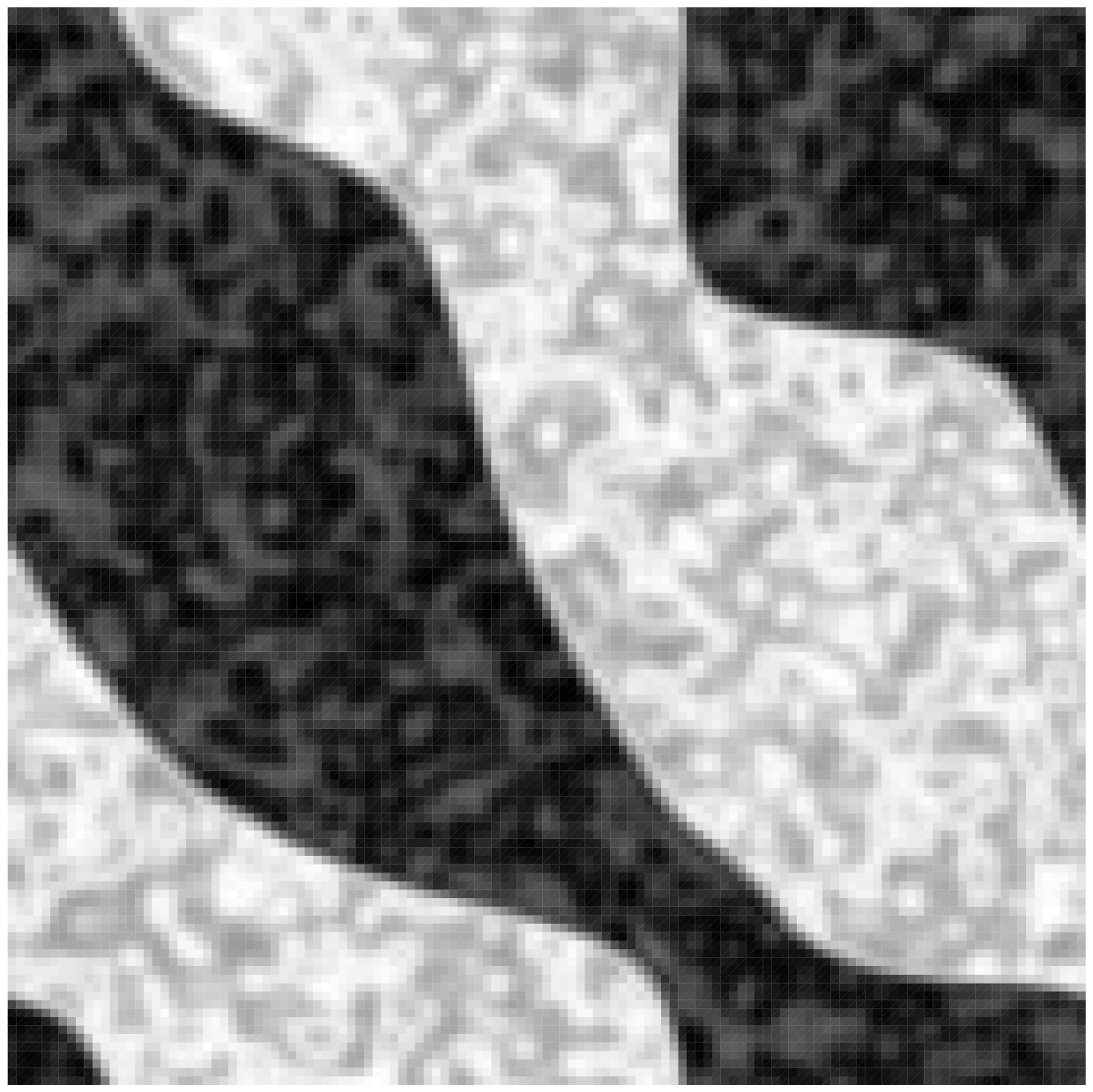}}}
\end{picture}
\end{center}
\begin{center}
\begin{picture}(200,200)(0,15)
\put(100,205){\makebox(0,0){\large (c)}}
\put(0,0){\makebox(200,200){\epsfxsize=180\unitlength
\epsffile{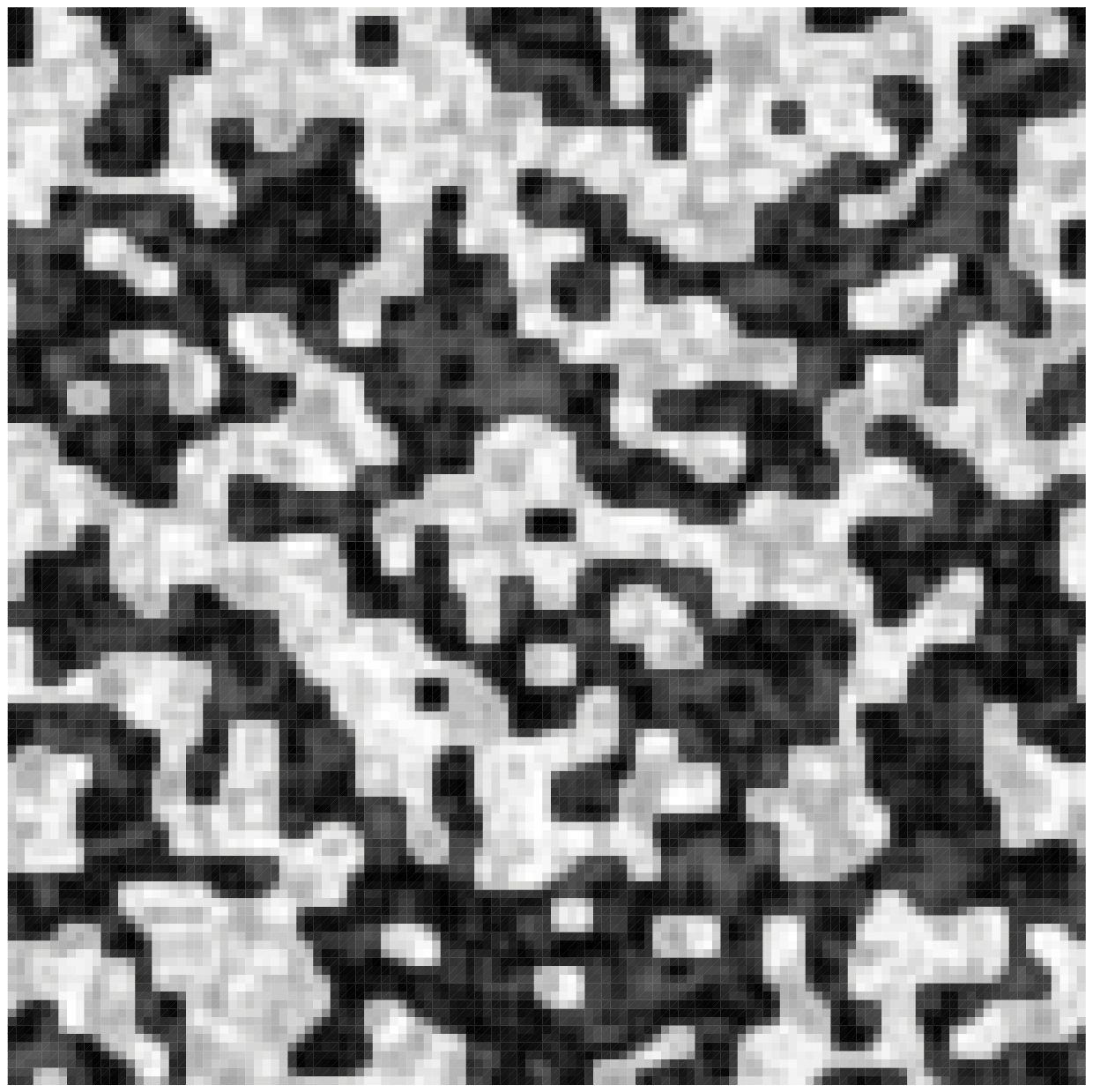}}}
\end{picture}
\hspace{10\unitlength}
\begin{picture}(200,200)(0,15)
\put(100,205){\makebox(0,0){\large (d)}}
\put(0,0){\makebox(200,200){\epsfxsize=180\unitlength
\epsffile{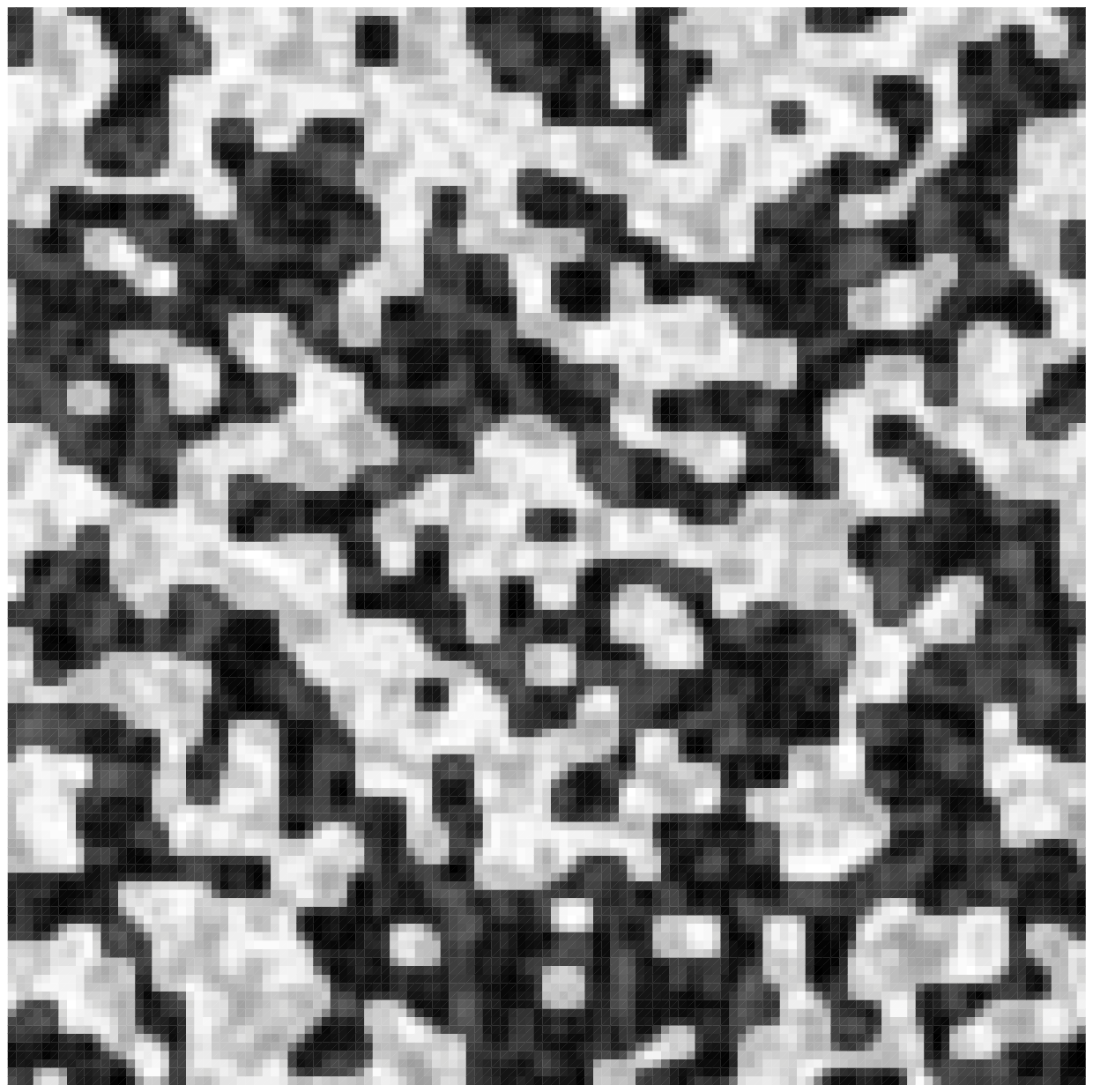}}}
\end{picture}
\end{center}
\caption{Snapshots of the $d=2$ CML with local map (\protect\ref{eq-mhmap}).
Lattice of $128^2$ sites, grey scale from $X=-1$ (white) to $X=1$ (black),
uncorrelated initial conditions.
(a,b): transient leading to complete ordering at $g=0.2>g_{\rm e}$, 
$t=100$ and 1000; 
(c,d): blocked state at $g=0.15<g_{\rm e}$, $t=1000$ and 2000.}
\label{fig-snap}
\end{figure}

\begin{figure}
\narrowtext
\unitlength = 0.0011\textwidth
\begin{picture}(200,200)(0,-10)
\put(0,0){\makebox(200,200){\epsfxsize=260\unitlength\epsffile{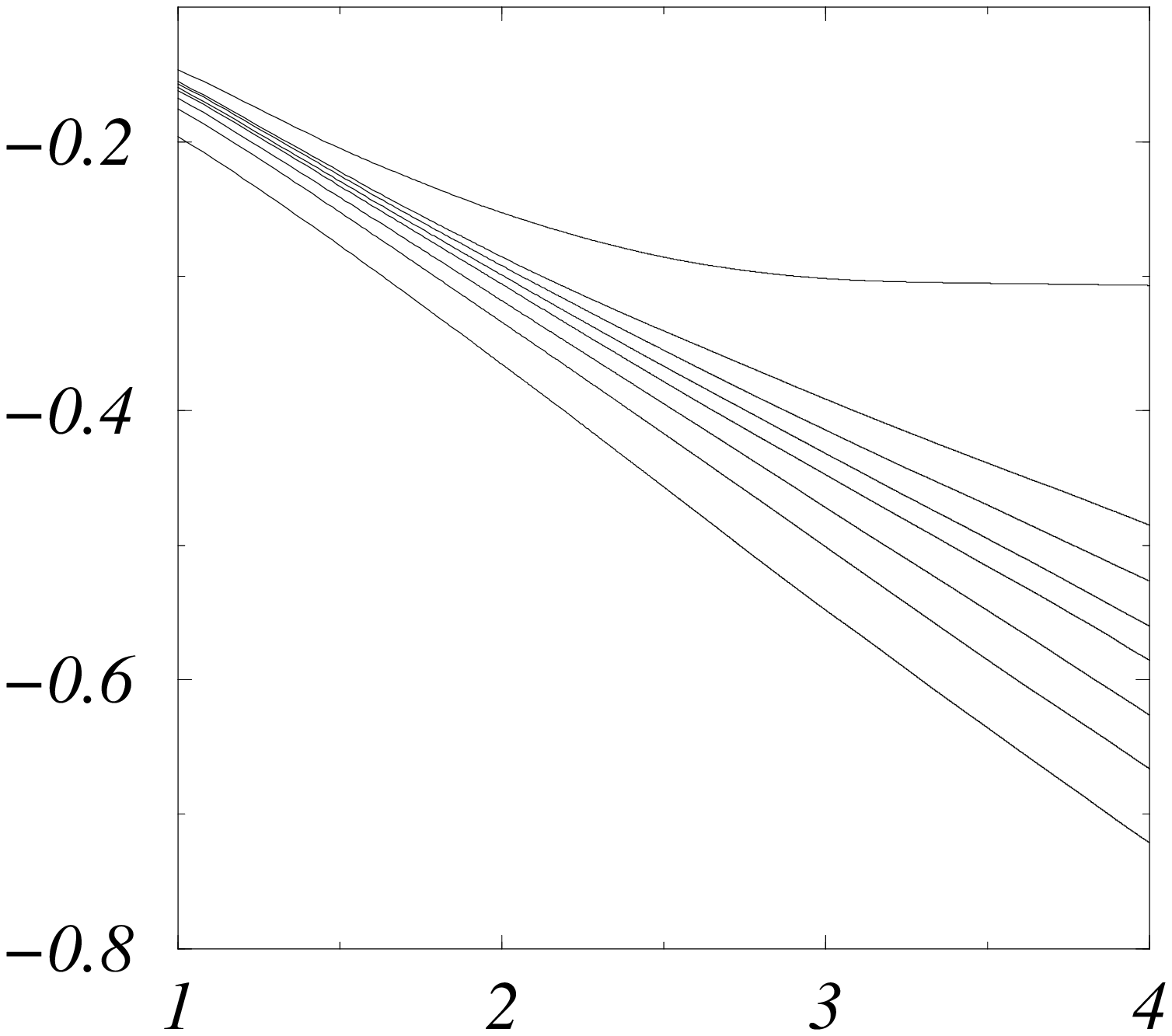}}}
\put(100,205){\makebox(0,0){(a)}}
\put(0,145){\makebox(0,0){$\log p$}}
\put(170,0){\makebox(0,0){$\log t$}}
\end{picture}
\hspace{10\unitlength}
\begin{picture}(200,200)(0,-10)
\put(0,0){\makebox(200,200){\epsfxsize=260\unitlength\epsffile{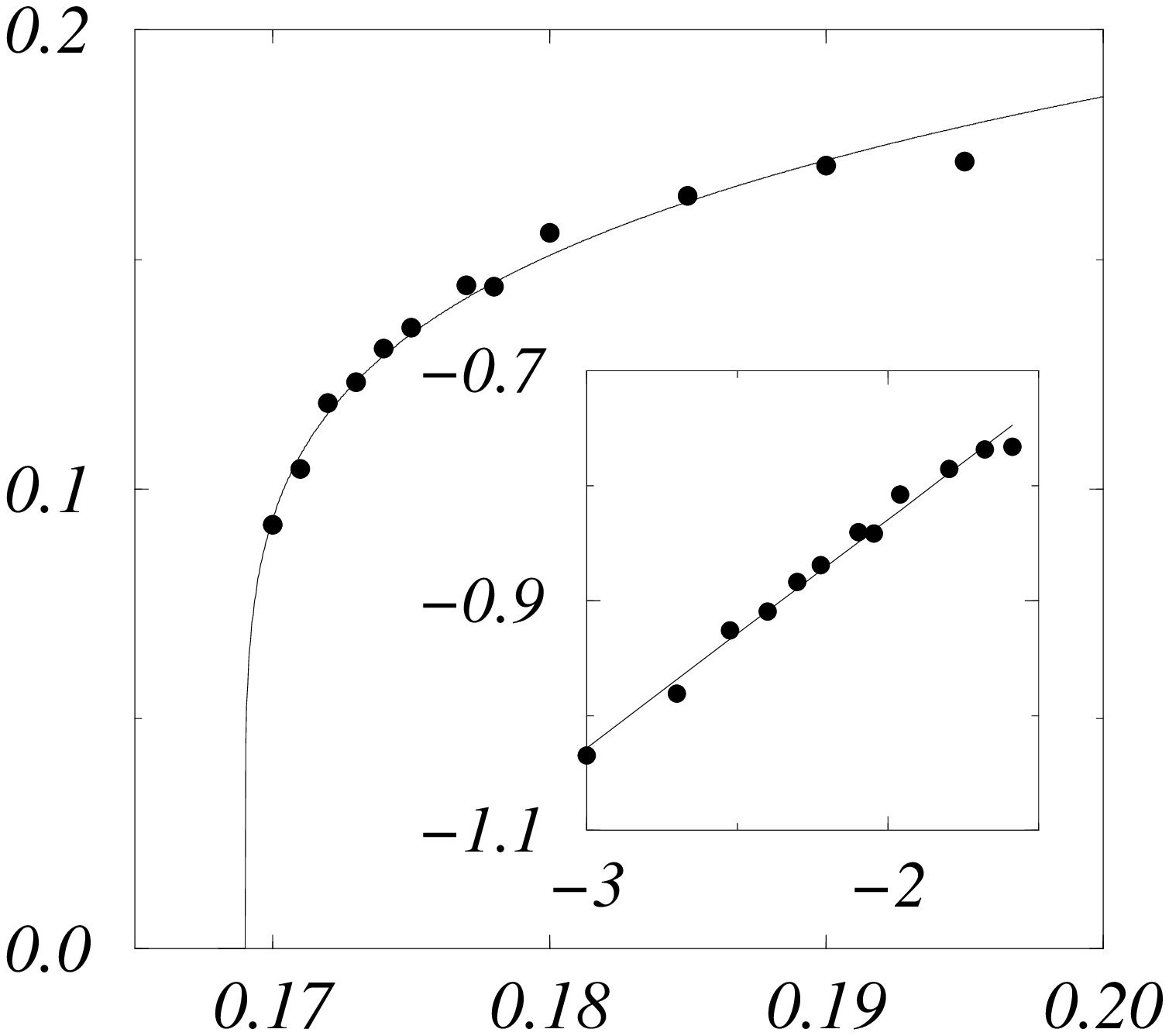}}}
\put(100,205){\makebox(0,0){(b)}}
\put(0,165){\makebox(0,0){$\theta$}}
\put(170,0){\makebox(0,0){$g$}}
\put(120,115){\makebox(0,0){$\scriptstyle\log \theta$}}
\put(150,60){\makebox(0,0){$\scriptstyle\log(g\!-\!g_{\rm e}\!)$}}
\end{picture}
\caption{Phase ordering in  the $d=2$ CML with local map 
(\protect\ref{eq-mhmap}) at $\mu=1.9$.
(a): algebraic decay of the persistence probability $p(t)$ for,
from top to bottom: $g=0.16$ ($< g_{\rm e}$: saturates to a finite level), 
0.17, 0.172, 0.174, 0.176, 0.18, 0.185, and 0.195.
(b): variation of persistence exponent $\theta$ with $g$. Solid line: fitting
Ansatz $\theta\sim (g-g_{\rm e})^w$ with $g_{\rm e}=0.169(1)$ and $w=0.20(3)$
Insert: $\log(\theta)$ vs. $\log(g-g_{\rm e})$.}
\label{fig-persis}
\end{figure}

For large $g$ values, complete 
phase ordering occurs (Fig.~\ref{fig-snap}a,b), 
and the system eventually reaches a regime in which
all sites are situated in one of the two intervals $I^\pm$.
For small $g$, initial conditions with sites in both intervals $I^\pm$
lead to spatially-blocked configurations 
where interfaces between clusters of
each phase are strictly pinned, while chaos is present within clusters 
(Fig.~\ref{fig-snap}c,d).~\cite{NOTE}

To study the phase ordering process efficiently, 
uncorrelated initial conditions were generated as follows:
exactly one half of the sites of a $d=2$ lattice were chosen at random
and assigned positive $X$ values drawn according to the invariant distribution
of $S_\mu$ on $I^+$, while the other sites were similarly assigned negative 
values. Large lattices with periodic boundary conditions were
used, and the persistence $p(t)$ was measured. 
Fig.~\ref{fig-persis}a shows the results of single runs for various values of 
$g$. For small $g$, $p(t)$ saturates at large times to strictly
positive values, while it decays algebraically, for large $g$, on square
lattices of linear size 2048 sites.
The associated persistence exponent 
$\theta$ varies continuously with $g$,
and its $g$-dependence is nicely accounted for by a functional form
$\theta \sim (g-g_{\rm e})^w$ with $g_{\rm e} \simeq 0.169(1)$ and 
$w\simeq 0.20(3)$ (Fig.~\ref{fig-persis}b).
We have, at this point, no theoretical justification of this fitting Ansatz.
At any rate, it provides an operational definition of $g_{\rm e}$ yielding
estimates consistent with those obtained using other, less accurate,
 methods \cite{TBP}.

\begin{figure}
\narrowtext
\unitlength = 0.0011\textwidth
\begin{picture}(200,200)(0,-10)
\put(0,0){\makebox(200,200){\epsfxsize=260\unitlength\epsffile{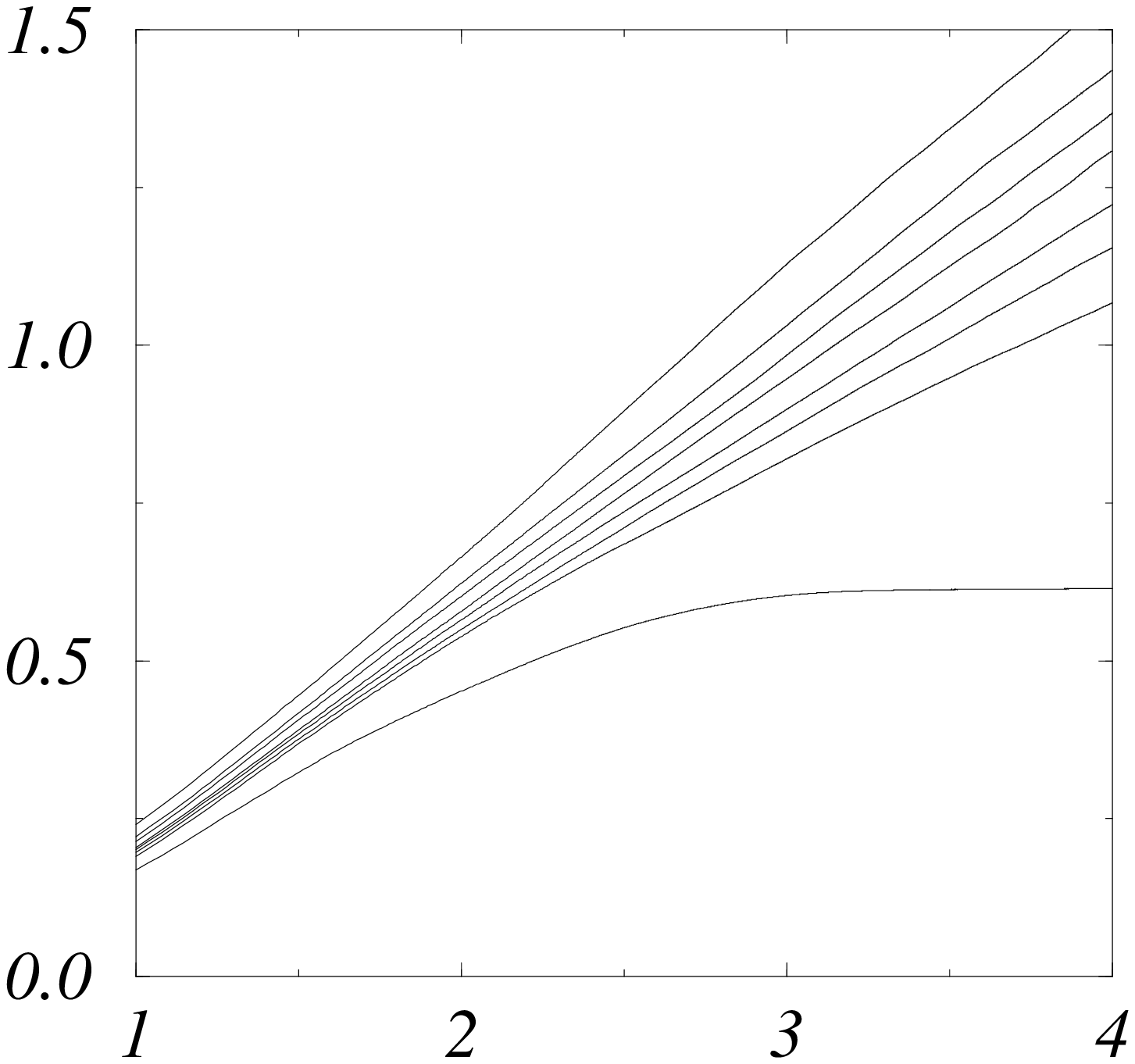}}}
\put(100,205){\makebox(0,0){(a)}}
\put(0,165){\makebox(0,0){$\log L$}}
\put(170,0){\makebox(0,0){$\log t$}}
\end{picture}
\hspace{10\unitlength}
\begin{picture}(200,200)(0,-10)
\put(0,0){\makebox(200,200){\epsfxsize=260\unitlength\epsffile{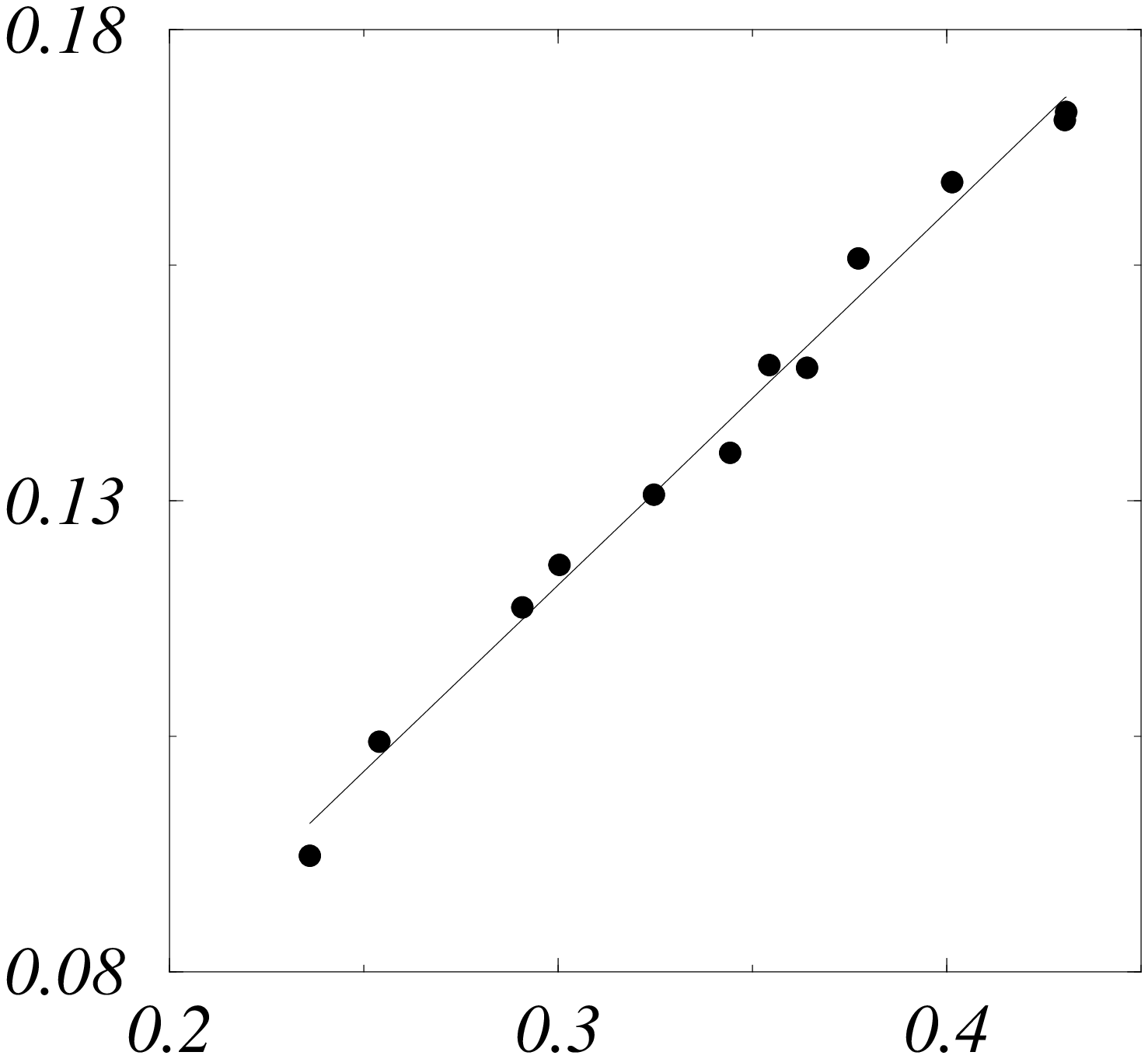}}}
\put(100,205){\makebox(0,0){(b)}}
\put(0,165){\makebox(0,0){$\theta$}}
\put(170,0){\makebox(0,0){$\phi$}}
\end{picture}
\caption{Same runs as in Fig.~\protect\ref{fig-persis}.
(a) $\log(L)$ vs $\log (t)$ (bottom curve: $g=0.16$, top curve $g=0.195$);
(b) $\theta(g)$ vs $\phi(g)$ for $g$ values between 0.17 and 0.20; 
the solid line is the linear fit $\theta \simeq 0.396\phi-0.002$.}
\label{fig-length}
\end{figure}

The origin of this unusual behavior of the persistence exponent is largely 
explained by the evolution of the spatial structures formed during 
phase ordering. Usually, one expects the coarsening to be described by
the algebraic growth of a single characteristic length $L(t) \sim t^\phi$
with $\phi=1/2$ for a non-conserved, scalar order parameter \cite{MODELA}.
In the CML studied above, the two-point correlation function 
$C(\vec x,t)=\langle \sigma_{{\vec r}+{\vec x}}^t \sigma_{{\vec r}}^t\rangle$
was measured during phase ordering \cite{NOTE1}.
Length $L(t)$ was then evaluated to be the width at mid-height 
($C(L(t),t)=1/2$), determined by interpolation.
This procedure was then validated 
by a collapse of all $C(||{\vec x}||/L(t),t)$ curves.
Surprisingly, while the scaling behavior of $L(t)$ is observed, exponent
$\phi$ departs from the expected $1/2$ value and varies continuously with $g$
(Fig.~\ref{fig-length}). Again, we find a law of the form 
$\phi \sim (g-g_{\rm e})^w$ to be an acceptable Ansatz 
of our numerical results.
The estimated values of  $g_{\rm e}$ and $w$ are consistent,
within numerical accuracy, with those found when fitting $\theta(g)$.
This is corroborated by studying directly $p(t)$ vs $L(t)$
(not shown), or by plotting $\theta$ vs $\phi$ 
which confirms that the two exponents are proportional to each other
 (Fig.~\ref{fig-length}d).
Remarkably, the ratio $\theta/\phi$ is found to have, 
within our numerical accuracy,  the
 $d=2$ TDGLE value: $\theta/\phi \simeq 0.40(2) \simeq 2\theta_{\rm GL}
\simeq 0.40$ \cite{PER-GL}. (We cannot, however, completely exclude the values
corresponding to the Ising model, or the diffusion equation, since
$\theta_{\rm Ising}\simeq 0.22$ \cite{PER-ISING}, and 
$\theta_{\rm Diff.}\simeq 0.19$ \cite{PER-GL}.)

\begin{figure}
\narrowtext
\unitlength = 0.0011\textwidth
\begin{picture}(200,200)(0,-10)
\put(0,0){\makebox(200,200){\epsfxsize=260\unitlength\epsffile{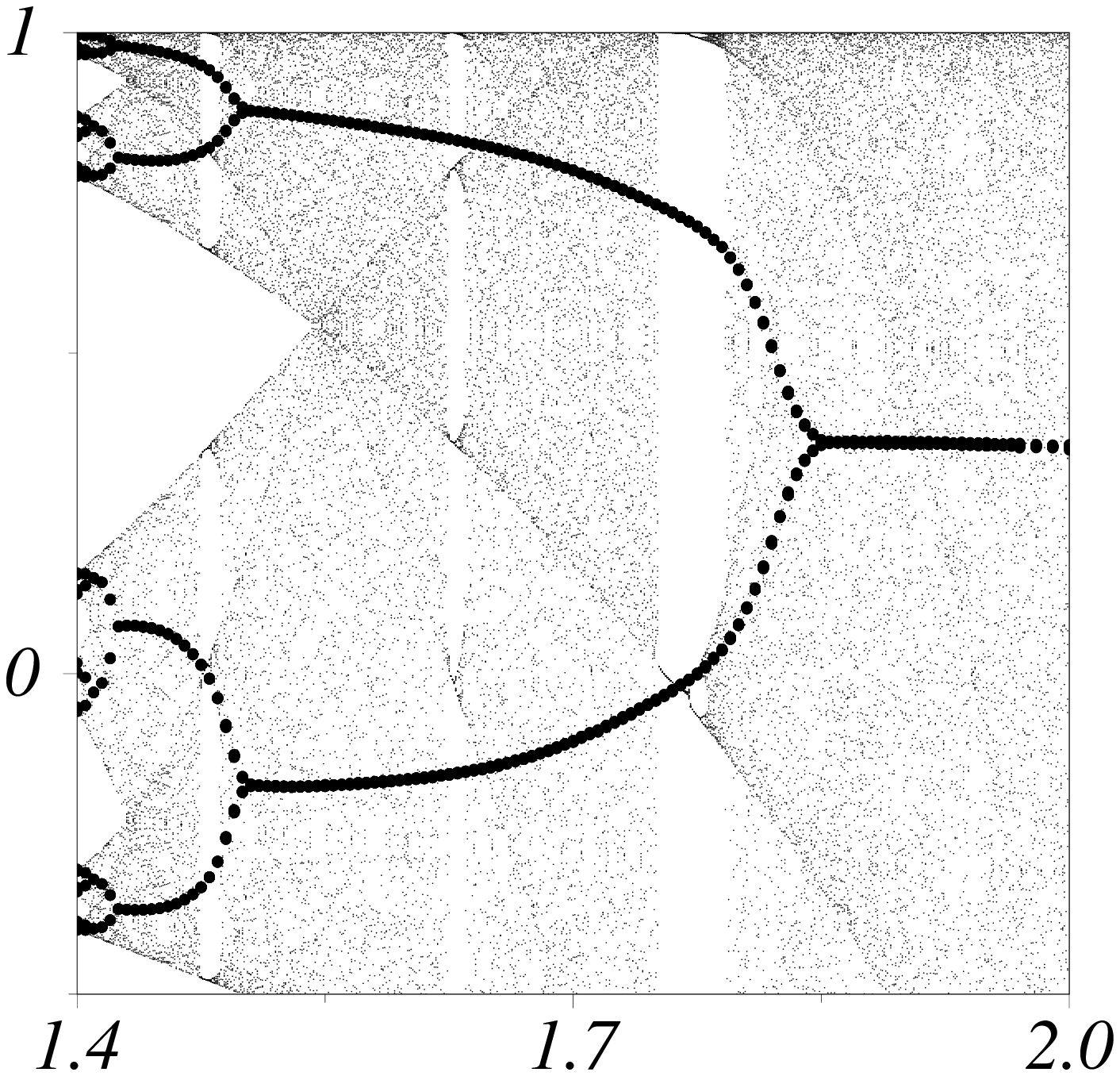}}}
\put(100,205){\makebox(0,0){(a)}}
\put(0,165){\makebox(0,0){$M^t$}}
\put(170,0){\makebox(0,0){$\mu$}}
\end{picture}
\hspace{10\unitlength}
\begin{picture}(200,200)(0,-10)
\put(0,0){\makebox(200,200){\epsfxsize=260\unitlength\epsffile{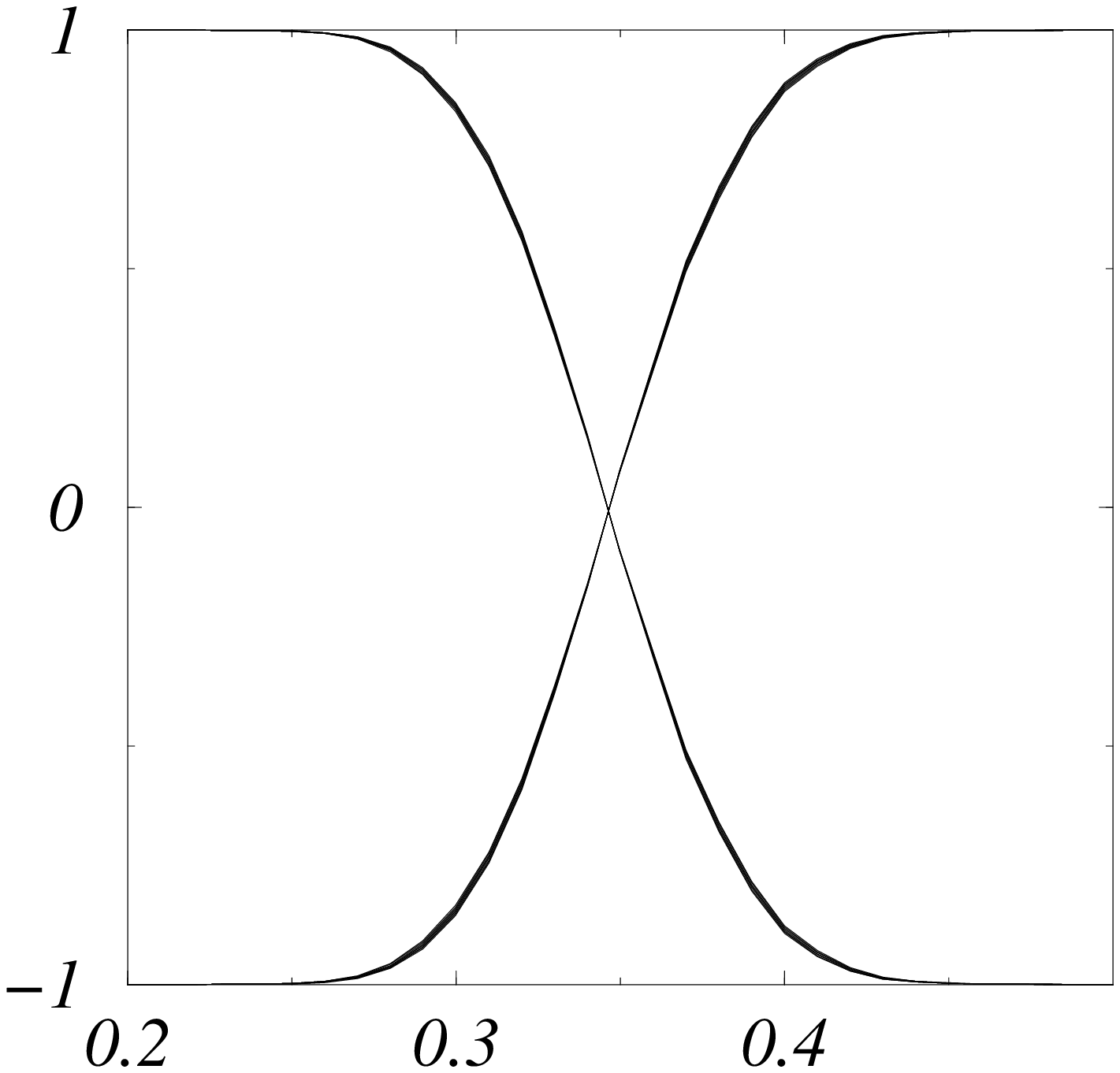}}}
\put(100,205){\makebox(0,0){(b)}}
\put(0,165){\makebox(0,0){$\langle{\bf \sigma}_{\vec r\,}\rangle^t$}}
\put(170,0){\makebox(0,0){$\rho$}}
\put(106,50){\makebox(0,0){$\rho^*$}}
\put(106,60){\vector(0,1){35}}
\end{picture}
\caption{$d=2$ lattice of coupled logistic maps for $g=0.2$. 
(a) bifurcation diagram
($\langle X_{\vec r}^t\rangle_{\vec r}$, large dots) superimposed on that
of single logistic map (small dots).
(b): $\langle \sigma_{\vec r}^t \rangle_{\vec r}$ vs $\rho$ plotted at
20 different timesteps between $t=500$ and $t=1000$ 
for $\mu^{\rm c}_2 < \mu=1.5 < \bar{\mu}_2$. This ``magnetization''
remains constant only for $\rho=\rho^*\simeq 0.3465$, whereas it reflects
the overall period-2 dynamics for all other $\rho$ values.
(c) $\theta(g)$ at $\mu=1.5$ and $\rho=\rho^*(g)$ with log-log fit 
($g_{\rm e}\simeq 0.101$, $w\simeq 0.06$).
(d) $\theta(g)$ vs $\phi(g)$; 
the solid line is the linear fit $\theta \simeq 0.40\phi-0.002$.}
\label{fig-log}
\end{figure}

The same analysis was also performed on CMLs with a non-symmetric, unimodal, 
local map $S_\mu$ of the form:
\begin{equation}
S_\mu(X) = 1-\mu |X|^{1+\varepsilon} \;\; {\rm with} \;\;\mu\in [0,2] \;,
\label{eq-unimap}
\end{equation}
in particular for $\varepsilon=0$ (tent map) and  $\varepsilon=1$
(logistic map). For $\mu\in [\mu_\infty, 2]$, this map shows 
$2^n$-band chaos and exhibits an inverse cascade of band-merging points 
$\bar{\mu}_n$ when $\mu\to\mu_\infty$.
In the strong-coupling limit, the corresponding  CMLs exhibit,
depending on $d$,
periodic or quasiperiodic NTCB  with a period equal to,
or a multiple of, that of the band-chaos of the local map 
\cite{CML-NTCB,CMLRG}. For $d=2$ and $3$, in particular, simple period-$2^n$
NTCB occurs, with an infinite cascade of phase transition points 
$\mu^{\rm c}_n$ distinct from the band-merging points 
(Fig.~\ref{fig-log}a). 
When period-2 NTCB occurs in the two-band chaotic region of the map
($\mu\in [ \bar{\mu}_2,\bar{\mu}_1 ]\approx[1.43,1.54]$),  
two-state spin variables $\sigma_{\vec r}^t \in \{-1,1\}$ can be defined,
but the asymmetry of the two bands hinders the generation of ``effectively''
uncorrelated initial conditions.
Indeed, an equal proportion of sites in each band quickly leads to
complete phase ordering and saturation of $p(t)$, even in the strong-coupling
regime. This happens because these initial conditions create, 
after a few timesteps, configurations with a fairly large unbalance
between the two phases.
Tuning the initial proportion $\rho$ of sites in, say, the band containing
$X=0$, one can minimize such effects. We determined the optimal proportion
$\rho^*$ defined as the value for which
the magnetization $\langle \sigma_{\vec r}^t \rangle$ remains constant
(Fig.~\ref{fig-log}b).
Clean scaling behavior of $L(t)$ and $p(t)$ is then observed with reasonable
system sizes, as with the symmetric local map (\ref{eq-mhmap}).
Varying the coupling strength $g$, exponents $\phi$ and 
$\theta$ show the same behavior as above, decreasing continuously to 
zero at $g_{\rm e}$. Fig.~\ref{fig-log}c shows the case of coupled logistic maps,
for which the Ansatz $\theta,\phi \sim (g-g_{\rm e})^w$ is, again, valid,
although not as good as in the case of local map (\ref{eq-mhmap}).
Note that the estimated value $w\simeq0.06(2)$ is different from that measured
for the CML with local map (\ref{eq-mhmap}), but 
$\theta/\phi\simeq 0.48(4)$ is still rather close to the TDGLE value
(Fig.~\ref{fig-log}d).

We now deal with the onset of more complex NTCB such as the 
period-$2^n$ cycles mentioned above for which the study of the phase ordering
in terms of two-state spin variables may not be legitimate.

Consider, for example, a CML with local map $S_\mu$ defined by
(\ref{eq-unimap}) in a 4-band chaotic regime 
($\mu\in [\bar{\mu}_3, \bar{\mu}_2]$) which exhibits period-4 NTCB.
The ``natural'' spin variables to study phase ordering take four values,
indexed by the 4-band chaotic cycle. However, these four bands can be 
grouped in two ``meta-bands'', since they arise from a band splitting 
bifurcation at $\bar{\mu}_2$, so that two-state spin variables can still be
defined. Accordingly, two limit coupling strengths can be defined:
$g_{\rm e}^1$, marking the onset of complete phase ordering between 
the two meta-bands,
and $g_{\rm e}^2$ for ordering from initial conditions  within one of
the meta-bands.
A priori, $g_{\rm e}^2 \ne g_{\rm e}^1$, and there might exist 
coupling strengths 
such that, e.g., pinned clusters exist within, but not between, 
the two meta-bands. The ``true'' onset of period-4 NTCB is then
given by $g \ge \max (g_{\rm e}^1, g_{\rm e}^2)$.
Similarly, for $\mu\in [\mu_\infty, \bar{\mu}_n]$, one can define
$n$ different $\mu$-dependent limit coupling strengths 
$g_{\rm e}^1, g_{\rm e}^2,  \ldots,  g_{\rm e}^n$,
with $n\to\infty$ as $\mu\to\mu_\infty$. 
Using our recent work on renormalisation group arguments for CMLs 
\cite{CMLRG}, one can show that the threshold values of 
this infinite hierarchy are related to each other. Here, we only describe
briefly these results, while a detailed derivation can be found in 
\cite{CMLRG}. 
The RG structure  of single map
(\ref{eq-unimap}) induces the conjugacy between
$({\bf \Delta}_g^m \circ {\bf S}_\mu)^2$ and 
${\bf \Delta}_g^{2m} \circ {\bf S}_{q(\mu)}$,
where ${\bf S}_{\mu}$ transforms each variable $X_{\vec r}$ by $S_{\mu}$,
${\bf \Delta}_g^m$ is the diffusive operator applied $m$ times, and 
$q(\mu)=\mu^2$ for coupled tent maps. This relation can be shown to imply that
$g_{\rm e}^2(\mu,m) = g_{\rm e}^1(q(\mu),2m)$. Furthermore, using the
fact that $g_{\rm e}^n(\mu,m)$ decreases with $m$, one can prove that
the maximum $g_{\rm e}$ for all $n$, $\mu$, and $m$ is
$g_{\rm e}^* = g_{\rm e}^1\left(\bar\mu_1,1\right)$.
Thus, whenever $g \ge g_{\rm e}^*$, complete
ordering occurs for all bands. 

The above results are  at odds with the behavior of
usual models studied in phase ordering problems \cite{COARSE}.
But in both cases presented here, the exponent ratio $\theta/\phi$
seems to take the value expected for the TDGLE model. 
This ``weak universality''
is reminiscent of similar results found recently at the Ising-like
critical points shown by the same models \cite{MARCQ}.
We note, moreover, that, when $g$ is increased, 
$\phi$  approaches $1/2$ and $\theta$ reaches values close to
$\theta_{\rm GL}$.
We believe that this tendancy 
is mostly due to the lattice effects becoming less
and less important (although strict pinning does not occur for $g>g_{\rm e}$).
We have shown recently \cite{CMLRG} that,
in the continuous-space  limit of CMLs, the weak coupling regime disappears
($g_{\rm e}\to 0$), together with any pinning effects. One can thus wonder
whether, in this limit, one recovers more ``conventional'' 
phase ordering dynamics. 

The continuous limit of CMLs
such as those defined by (\ref{eq-cml}-\ref{eq-mhmap}) is reached when
applying the coupling step of the dynamics more and more times per iteration,
i.e. when taking the $m\to\infty$ limit of ${\bf \Delta}_g^m \circ 
{\bf S}_{\mu}$. In this limit,
${\bf \Delta}_g^m$ converges to a universal Gaussian 
kernel ${\bf \Delta}_{\lambda}^\infty = \exp (\frac{\lambda^2}{2}\nabla^2)$
with a coupling range $\lambda=\sqrt{2gm} \| {\vec e} \|$ where 
$\|{\vec e}\|$ is the lattice spacing, which can thus be chosen to scale
like $1/\sqrt{m}$ so as to keep $\lambda$ constant.
We investigated the phase ordering properties of these CMLs 
with the symmetric local map (\ref{eq-mhmap})
for increasing values of $m$. At a qualitative level, the
scaling behavior of $L(t)$ and $p(t)$ is observed at all $m$ values.
Quantitatively, exponents $\theta$ and $\phi$ vary with
$m$ at fixed $g$. Increasing $m$, $\phi$ seems to 
converge to $1/2$, while $\theta \to \theta_{\rm GL}$: for $m=1$ to 3,
we find $\phi=0.467$, 0.479, 0.505, and $\theta=0.174$, 0.184, 0.196,
from single runs on lattices of linear size 4096 sites.

Our work provides a quantitative method for determining the onset of
NTCB in chaotic coupled map lattices. It also
reveals that the phase-ordering properties of 
multiphase, chaotic CMLs are different from those of most models studied
traditionally. 
More work is needed, especially at the analytical level, 
to clarify the origin of the non-universality
observed and put our numerical results on firmer ground, 
since we cannot completely
exclude a very slow, unobservable, crossover
of the scaling behavior observed to that of a more traditional model.
Different approaches can be suggested.

A continuous variation of the scaling exponent $\phi$
for the characteristic length of domains is not usually observed, but 
(at least) two 
exceptions are known. One is the case of coarsening from
initial conditions with built-in long-range correlations
\cite{CORREL-IC}, but then the persistence probability $p(t)$
does {\it not} decrease algebraically with time \cite{NAKA}. 
Another situation of possible relevance is the case of phase-ordering
with an order-parameter-dependent mobility \cite{EMM}, for which, 
unfortunately, the behavior of the persistence is not known.
At any rate, the recovery of the ``normal''
scaling properties of the TDGLE in the space-continuous limit
suggests that lattice effects are ultimately responsible for the 
non-trivial scaling properties recorded in discrete systems.
This calls for
a detailed study of interface dynamics in order to assess the effective
role of discretization and anisotropy.

Finally, we believe our results are general and that similar behavior
should be found 
in experiments on phase-ordering of pattern-forming systems, such as, e.g.,
electro-hydrodynamical convection in liquid crystals, or
Rayleigh-B\'enard convection \cite{MCC}.

We thank Ivan Dornic for many fruitful discussions and his keen interest
in our work.

\end{multicols}

\begin{references}

\bibitem{GEN-NTCB} H. Chat\'e, Int. J. Mod. Phys. B {\bf 12}, 299 (1998).

\bibitem{CA-NTCB} J.~A.~C.~Gallas et al., Physica A {\bf 180}, 19 (1992); 
J.~Hemmingsson, Physica A {\bf 183}, 255 (1992); 
H.~Chat\'e and P.~Manneville, Europhys. Lett. {\bf 14}, 409 (1991);
 H. Chat\'e, L.-H. Tang, and G. Grinstein, 
Phys. Rev. Lett. {\bf 74}, 912 (1995).

\bibitem{CML-NTCB} H.~Chat\'e and P.~Manneville, Prog. Theor. Phys.
{\bf 87}, 1 (1992); Europhys. Lett. {\bf 17}, 291 (1992); H. Chat\'e
and J. Losson, Physica D {\bf 103}, 51 (1997).

\bibitem{ANTI-MULTI} R.S. MacKay and T.A. S\'epulchre, Physica D {\bf 82},
243 (1995); T.A. S\'epulchre and R.S. MacKay, Nonlinearity {\bf 10}, 679
(1997); S. Aubry, Physica D {\bf 103}, 201 (1997).

\bibitem{PERSIS} See, e.g.: 
I. Dornic and C. Godr\`eche, J. Phys. A {\bf 31}, 5413 (1998),
and references therein.

\bibitem{PER-ISING} D. Stauffer, J. Phys. A {\bf 27}, 5029 (1994).

\bibitem{MODELA} P.C. Hohenberg and B.I. Halperin, Rev. Mod. Phys.
{\bf 49}, 436 (1977).

\bibitem{CMLRG} A. Lema\^{\i}tre and H. Chat\'e, Phys. Rev. Lett. {\bf 80},
5528 (1998); preprint, 1998.

\bibitem{MH} J.~Miller and D.A.~Huse, Phys. Rev. E {\bf 48},  2528  (1993).

\bibitem{NOTE} In CMLs, pinning can be strict (the 
``effective noise'' arising from local chaos is bounded). It is {\it local}, 
and thus no finite-size effect is observed for large-enough systems.

\bibitem{TBP} A. Lema\^{\i}tre, Ph.D. thesis, Ecole Polytechnique, 1998.

\bibitem{NOTE1} The same analysis was also performed using the
continuous variables $X_{\vec r}$, yielding similar, albeit noisier, results.

\bibitem{PER-GL} S. Cueille and C. Sire, ``Block persistence'', 
cond-mat/9803014; S. H. Cornell, private communication.

\bibitem{PER-ISING} D. Stauffer, J. Phys. A {\bf 27}, 5029 (1994).

\bibitem{COARSE} A.J. Bray, Adv. Phys. {\bf 43}, 357 (1994).

\bibitem{MARCQ} P. Marcq, H. Chat\'e, and P. Manneville, 
Phys. Rev. Lett. {\bf 77}, 4003 (1996); Phys. Rev. E {\bf 55}, 2488 (1997).

\bibitem{CORREL-IC} B. Derrida, C. Godr\`eche, and I. Yekutieli, Phys.
Rev. A {\bf 44}, 6241 (1991).

\bibitem{NAKA} H. Nakanishi, H. Chat\'e, and I. Dornic, unpublished.

\bibitem{EMM} C.L. Emmott and A.J. Bray, preprint cond-mat/9808308.

\bibitem{MCC} M.C. Cross and D.I. Meiron, Phys. Rev. Lett. {\bf 75}, 2152
(1995).

\end{references}
\end{document}